\documentclass[12pt]{article}
\usepackage{graphics,epsfig}
\usepackage[english]{babel}
\newcommand{\cinst}[2]{$^{\mathrm{#1}}$~#2\par}
\newcommand{\crefi}[1]{$^{\mathrm{#1}}$}

\begin{document}


\thispagestyle{empty}
\begingroup



\begin{center}

{\Large{\bf The characteristics of neutrinonuclear reactions \\
\vspace{0.3cm}

at $E_{\nu}$ = 1 - 3 GeV}}

\end{center}

\vspace{1.cm}

\begin{center}
{\large SKAT Collaboration}

 N.M.~Agababyan\crefi{1}, V.V.~Ammosov\crefi{2},
 M.~Atayan\crefi{3},\\
 N.~Grigoryan\crefi{3}, H.~Gulkanyan\crefi{3},
 A.A.~Ivanilov\crefi{2},\\ Zh.~Karamyan\crefi{3},
V.A.~Korotkov\crefi{2}

\setlength{\parskip}{0mm}
\small

\vspace{1.cm} \cinst{1}{Joint Institute for Nuclear Research,
Dubna, Russia} \cinst{2}{Institute for High Energy Physics,
Protvino, Russia} \cinst{3}{Yerevan Physics Institute, Armenia}
\end{center}
\vspace{100mm}

{\centerline{\bf YEREVAN  2004}}

\newpage
\thispagestyle{empty}
\begin{abstract}
\noindent For the first time, the characteristics of the
charged current neutrinonuclear interactions
are investigated at $E_{\nu}$ = 1 - 3 GeV
using the data obtained with SKAT propane-freon bubble chamber
irradiated in the neutrino beam at Serpukhov accelerator.
The $E_{\nu}$ - dependence of the mean multiplicities of
different types of secondary particles and their multiplicity,
momentum and angular distributions are measured.
\end{abstract}

\newpage
\setcounter{page}{1}
\section{Introduction}

The experimental data on the characteristics of neutrinonuclear
reactions at $E_{\nu} < $3 GeV are of interest, in particular, in
view of current and anticipated experiments on the neutrino
oscillations (see, e.g., \cite{ref1}). The detailed data are
available only in the low energy range of $E_{\nu} < $1 GeV, where
the quasielastic reaction ${\nu}n\rightarrow {\mu}^- p$ on the
nuclear neutrons dominates \cite{ref2}. The aim of this work is to
obtain, for the first time, detailed experimental data on
neutrinonuclear reactions in the range of 1$< E_{\nu} < $3 GeV.
Section 2 is devoted to the experimental procedure of the
selection of the neutrino interaction events and the
reconstruction of the neutrino energy. The experimental results
are presented in Section 3 and summarized in Section 4.

\section{Experimental procedure}

{\it a) The event selection criteria}

\noindent The experiment was performed with SKAT bubble chamber
\cite{ref3}, exposed to a wideband neutrino beam obtained with a
70 GeV primary protons from the Serpukhov accelerator. The chamber
was filled with a propan-freon mixture containing 87 vol\% propane
($C_3H_8$) and 13 vol\% freon ($CF_3Br$) with the percentage of
nuclei H:C:F:Br = 67.9:26.8:4.0:1.3 \%. The nuclear interaction
and radiation lengths of the mixture were ${\lambda}_I$ = 149 cm
and $X_0$ = 50 cm. The volume of the chamber was 6.5 m$^3$, the
chosen fiducal volume 1.73 m$^3$. A 20 kG uniform magnetic field
was provided within the operating chamber volume. \\ Charged
current interactions, containing a negative muon were selected. A
muon was identified as a particle that possessed the highest
transverse momentum among negative particles and did not suffer a
secondary interaction in the chamber. Other negatively charged
particles were considered to be $\pi^-$ mesons. The overwhelming
part of protons with momentum below 0.6 GeV$/c$ and a fraction of
those with momentum up to 0.85 GeV$/c$ were identified by their
stopping in the chamber. Other positively charged particles were
assumed to be ${\pi}^+$ mesons, except for cases explained below.
\\ In order to reduce the uncertainty in reconstructing the
neutrino energy $E_{\nu}$, the events were selected for which
errors in measuring the momenta of all charged secondaries and
converted photons were less than 27\% and 100\%, respectively. The
mean relative error $< \Delta p/p >$ in the momentum measurement
for muons, protons, pions and gammas was, respectively, 4\%, 6\%,
10\% and 21\%. Each selected event was assigned a weight that took
into account loss of events. The mean weight of the sample used in
the present study was 1.18.\\ To diminish the background induced
by neutral and charged particles, the following requirements were
imposed on selected events:  \\
    - The net charge of secondary hadrons is positive. \\
    - The visual energy of neutrino $E_{\nu}^{vis} = E_{\mu} + {\nu}_{vis} >$ 0.6
    GeV, where $E_{\mu}$ is the muon energy, ${\nu}_{vis}$ is the
    visual energy transferred to hadrons (including decay photons and
    neutral strange particles, if any). \\
    - The value of the summary momentum $\overrightarrow{p}$ of
    detected secondary particles was $\mid \overrightarrow{p} \mid
    >$ 0.7 GeV$/c$, while its longitudinal component $p_L$ (with
    respect to the neutrino direction) exceeded the transverse
    component $p_T$ (the latter coinciding with the summary transverse
    momentum of undetected secondary particles). \\
    - The minimal neutrino energy defined from the kinematics of
    the quasielastic reaction ${\nu}n\rightarrow {\mu}^- p$,
    $E_{\nu}^{el} = (m_{N}E_{\mu} - 0.5 {m_{\mu}}^2) / (m_N - E_{\mu}
    + p_{\mu} \cos {\vartheta}_{\mu})$, did not exceed more than
    by 30\% \cite{ref2} the upper limit (3 GeV) of the considered
    range of $E_{\nu}$: $E_{\nu}^{el} <$ 3.9 GeV. Here $m_N$ and
    $m_{\mu}$ are the proton and muon masses, $p_{\mu}$ and
    ${\vartheta}_{\mu}$ are the muon momentum and angle with
    respect to the neutrino. \\ The distribution on
    $E_{\nu}^{vis}$ for selected events in the range 0.6 $<
    E_{\nu}^{vis} <$ 4 GeV is shown in Fig. 1a. The measured value
    of $E_{\nu}^{vis}$ is, in general, the minimal value of the
    true neutrino energy, and hence to be corrected for the
    contribution from undetected gammas and neutrons. An exception
    are the events satisfying the kinematics of exclusive
    reactions without secondary neutral particles (see the next
    subsection). For a part of these events, $E_{\nu}$ can be even
    smaller than $E_{\nu}^{vis}$ provided that the proton
    hypothesis for an unidentified positive particle matches
    better the kinematics of an exclusive channel, than the
    ${\pi}^+$ hypothesis. \\

\noindent {\it b) The identification of exclusive channel}

    \noindent In order to avoid a unnecessary energy correction for
    undetected secondaries, each event was analysed for its
    belonging to the following exclusive channels without neutral
    secondaries:

\begin{eqnarray}
\nu n &\rightarrow& {\mu}^{-} p \qquad
\\
\nu p &\rightarrow& {\mu}^{-} p \,\, \pi^+  \qquad
\\
\nu n &\rightarrow& {\mu}^{-} p \, \, \pi^+ \pi^-   \qquad
\end{eqnarray}

\noindent with an arbitrary number of accompanying identified
protons (originating from the nuclear cascading). In (1)-(3), p
($\pi^+$) stands for an identified proton ($\pi^+$ meson) or a
positively charged particle to be identified as a proton or
$\pi^+$ meson from the kinematical fit. \\ Figs. 2a and 2b show
the distributions on the azimutal angle $\phi$ between vectors $
\overrightarrow{p}_{\mu}$ and $\overrightarrow{p}_h =
\overrightarrow{p}_ - \overrightarrow{p}_{\mu}$ in the transverse
plane and on the squared transverse momentum $p^2_T$ for events
topologically consistent with reactions (1)-(3). The observed
peaks near $\phi \sim 180^{\circ}$ and $p^2_T \sim$ 0 indicate on
a significant fraction of events satisfying the kinematics of
reactions (1)-(3). For further analysis the events with $\phi >
155^{\circ}$ and $p_T^2 <$ 0.15 (GeV/$c)^2$ were selected, the
chosen boundaries roughly corresponding to violation of the
kinematics of reactions (1)-(3) due to the Fermi-motion of the
target nucleon and the measurement errors. For these events, the
distribution on the dimensionless variable $r_L = (E_{\nu}^{vis} -
p_L)/E_{\nu}^{vis}$, characterizing the longitudinal
momentum-energy disbalance of the reaction, is plotted in Fig. 2c.
As it is seen, the distribution is strongly peaked at $r_L
\approx$ 0. The events with $ \mid r_L \mid <$ 0.15 were
assumed to be belonging to the exclusive reactions (1)-(3). The
number of these events is 126 (the weighted number 140.2), the
distribution of which on $E_{\nu}$ is plotted in Fig. 2d and,
separately for channels (1), (2) and (3), in Fig. 2e.
The curves in this figure are obtained by approximation of available
experimental data for $\nu p$ and $\nu n$ interactions \cite{ref4}
(taking into account the proton and neutron contents of the
target nuclei) convoluted with the expected neutrino flux
spectrum on the bubble chamber SKAT \cite{ref5}. The normalization
of these curves is chosen to fit our data on the main exclusive
channel (1). As a result, the predictions for channels (2) and (3)
overestimate the data, probably due to the intranuclear interaction
effects for pions produced in the primary  $\nu N$ interaction
(see Section 3 below). \\
The distribution on the invariant mass $W$ of the hadronic
system for exclusive channels (1) and (2) is shown in Fig. 2f.
The clear peak at $W \approx m_N$,
with a typical width $\sim$ 0.15 GeV (caused by
Fermi-motion and measurement errors), corresponds to the
quasielastic reaction (1). The $W$- distribution for the
reaction (2) reflects probably the production of the
${\Delta}^{++}$ states. \\

\noindent {\it c) The reconstruction of the neutrino spectrum}

\noindent Only a small part (about 28\%) of the total event sample
can be identified as exclusive channels (1)-(3). The rest part
contains undetected secondaries, for which a correction has to be
introduced to estimate $E_{\nu}$. The mean value $\nu$ of the
corrected energy transferred to hadrons (at given narrow intervals
of ${\nu}_{vis}$) is approximated as $\nu = a + b{\nu}_{vis}$,
where the coefficients $a$ and $b$ are found by means of the
procedure applied in \cite{ref6}: $a = 0.15 \pm 0.24$ GeV, $b =
1.07 \pm 0.05$ at $E_{\nu}^{vis} >$ 2 GeV and $a = 0.5 \pm
0.3$ GeV, $b = 1.5 \pm 0.4$ at $E_{\nu}^{vis} <$ 2 GeV. The
reconstructed spectrum of $E_{\nu}$ in the range 1 - 3 GeV is
presented in Fig. 1b, while Fig. 1c shows the spectrum of the beam
neutrino evaluated using the total cross sections \cite{ref7} of
reactions $\nu p \rightarrow {\mu}^{-} X$ and $\nu n \rightarrow
{\mu}^{-} X$ and taking into account the proton and neutron
content of the target nuclei. As it is seen, the latter spectrum
is consistent with the expected neutrino flux spectrum (at
$E_{\nu} >$ 1.5 GeV) on the bubble chamber SKAT \cite{ref5}.

\section{Experimental results}

{\it a) The  $E_{\nu}$- dependence of the mean characteristics of
produced particles}

\noindent Fig. 3 presents the $E_{\nu}$ - dependence of mean
multiplicities of secondary particles: protons, ${\pi}^+$ mesons,
${\pi}^-$ mesons, charged pions, $\gamma$- quanta, $\pi^0$ mesons
and neutral strange particles ($V^0$). The mean value of
$<n_{{\pi}^0}>$ is estimated as $<n_{{\pi}^0}> = 0.5 <n_{\gamma}>
\bar {w_{\gamma}}$, where $\bar {w_{\gamma}}$ is the mean weight
of detected $\gamma$'s defined by their conversion probability in
the chamber. As it is seen, all multiplicities increase with
$E_{\nu}$, except the mean multiplicity of identified protons,
probably due to the fact that, with increasing $E_{\nu}$, a larger
fraction of recoil protons acquires momenta $p_p >$ 0.6 GeV$/c$
and hence can be detected as a unidentified positive particle. The
contribution of these protons is partly taken into account in Fig.
3a, where the full circles stand for $<n_p>$ including also
'kinematically' identified protons from reactions (1)-(3). The
latter will be included below in the all data concerning protons.
The solid curves in Figs. 3b, 3c and 3f are the approximation of
the experimental data \cite{ref4} on the mean multiplicities of
${\pi}^+$, ${\pi}^-$ and ${\pi}^0$ mesons in $\nu N$ interactions.
It is seen, that the values of $<n_{{\pi}^+}>$ at $E_{\nu}$ = 2.3
- 3 GeV  exceed those in $\nu N$ interactions, probably due to a
significant ($\sim$ 30\%) contamination from the unidentified
protons. On the other hand, the values of $<n_{{\pi}^+}>$ at
$E_{\nu}$ = 1 - 1.9 GeV and $<n_{{\pi}^-}>$ and $<n_{{\pi}^0}>$ in
the whole $E_{\nu}$ range are reduced as compared to those in $\nu
N$ interactions. This reduction can be, at least partly,
attributed to the absorption of pions on a pair of nucleons within
the nucleus, $\pi (NN) \rightarrow (NN')$. The dashed curves in
Figs. 3b, 3c and 3f are the results of calculations incorporating
this process (see for details \cite{ref2} and references therein).
As it is seen, the predictions are in reasonably agreement with
the data on $<n_{{\pi}^+}>$ at $E_{\nu}$ = 1 - 1.9 GeV, as well as
on $<n_{{\pi}^-}>$ and $<n_{{\pi}^0}>$ in the whole energy
interval, especially if one takes into account the uncertainties
(reaching $\pm$ 20\%) in the approximation of the experimental
data on the mean multiplicity of pions in $\nu N$ interactions. \\
\noindent The  $E_{\nu}$ - dependence of the mean momentum of
different types of particles is presented in Fig. 4. This
dependence is more expressed for charged pions and especially for
muons, being rather moderate for protons and gammas. Fig. 5 shows
the $E_{\nu}$ - dependence of the mean fraction $f_i$ of the
neutrino energy transferred to the muon ($f_{\mu} =
E_{\mu}/E_{\nu}$), protons ($f_p = \sum{T_p}/E_{\nu}$), charged
pions ($f_{\pi} = \sum{E_{\pi}}/E_{\nu}$), detected gammas
($f_{\gamma} = \sum{E_{\gamma}}/E_{\nu}$) and undetected neutral
particles ($f_{miss} = ({\nu} - {\nu}_{vis}){/E_{\nu}}$). It is
seen, that $f_{\mu}$ depends on $E_{\nu}$ rather weekly, being
enclosed between 0.41 $<$ $f_{\mu}$ $<$ 0.47. The proton
contribution smoothly decreases from $f_p$ = 0.13 to 0.05, while
that for charged pions increases from $f_{\pi}$ = 0.12 to 0.32.
The contribution of the detected gammas is very small, amounting
$f_{\gamma}$ = 0.01 - 0.04, and increases with $E_{\nu}$. 
The energy fraction carried by the undetected neutral particles 
decreases from
$f_{miss}$ $\sim$ 0.3 up to $\sim$ 0.1 with increasing $E_{\nu}$.
\\ The mean values of the invariant mass $W$ of the hadronic
system and the squared transfer momentum $Q^2$ are plotted in
Fig.~6. A significant rise of $<Q^2>$ is observed, while the
$E_{\nu}$- dependence of $<W>$ is rather weak. \\

\noindent {\it b) The multiplicity, momentum and angular distributions
of produced particles}

\noindent Figs.~7a - 7f present the multiplicity and momentum
distributions for protons, as well as separately for those
produced in the forward ($cos{\theta_p} >$ 0) and backward
($cos{\theta_p} < $0) directions with respect to the neutrino
beam. These distributions, as well as the proton angular
distribution (Fig.~7g), are presented for the whole range of
$E_{\nu}$ = 1 - 3 GeV, because the characteristics of protons
were found to not depend noticeably on $E_{\nu}$.
The multiplicity distributions, shown in Figs.~7a - 7c,
are compared with the poissonian distributions calculated at
the measured values of $<n_p> = 1.17 \pm 0.05$,
$<n_p> = 0.83 \pm 0.04$ (at $cos{\theta_p} >$ 0) and
 $<n_p> = 0.33 \pm 0.02$ (at $cos{\theta_p} <$ 0). As it is seen, the
tail of experimental distributions deviates from the poissonian
one. The momentum distributions (Figs.~7d - 7f) can be roughly
approximated by the dependence $\sim$ $ exp(-{\alpha} p^2)$ with
$\alpha = 3.5 \pm 0.2$  (at all $cos{\theta_p}$), 2.8 $\pm$ 0.2
(at $cos{\theta_p} >$ 0) and 4.6 $\pm$ 0.4 (GeV/$c)^{-2}$ (at
$cos{\theta_p} <$ 0). The distribution on  $cos{\theta_p}$
(Fig.~7g) exhibits a linear dependence $\sim$ 1 + $\beta
cos{\theta_p}$ with a slope parameter $\beta$ = 0.82 $\pm$ 0.05.
\\ Figs.~8 and 9 show the momentum and angular distributions of
muons and charged pions in two intervals of $E_{\nu}$ = 1 - 2.5
and 2.5 - 3 GeV with approximately equal statistics. The momentum
distribution of muons (Figs.~8a and 8b) is peaked at {$p_\mu$}
$\sim$ 0.5 GeV/$c$ and falls steeply with increasing {$p_\mu$} for
the range $E_{\nu}$ = 1 - 2.5 GeV, being, however, much flatter
for the range $E_{\nu}$ = 2.5 - 3 GeV. The angular distribution of
muons is similar for both $E_{\nu}$ ranges, being strongly peaked
at $cos{\theta_\mu}$ $\sim$ 1. The momentum distribution of pions
is also peaked at  {$p_\pi$} $\sim$ 0.5 GeV/$c$ and falls (more
steeply for the range $E_{\nu}$ = 1 - 2.5 GeV) with increasing
{$p_\pi$}. Their angular distribution is peaked at
$cos{\theta_\mu}$ $\sim$ 1 (more strongly for the range $E_{\nu}$
= 2.5 - 3 GeV). \\ The momentum and angular distributions of
$\gamma$- quanta, both corrected and not corrected for their
detection efficiency, are shown in Fig.~10, but now, due to the
lack of their statistics, for the whole range of $E_{\nu}$ = 1 - 3
GeV. The momentum distribution has a maximum at {$p_\gamma$}
$\sim$ 0.15 GeV/$c$ (more than twice smaller than that for charged
pions), while the angular distribution is somewhat shifted toward
larger angles as compared to that for charged pions.\\ Finally,
the distributions on $W$ and $Q^2$ are plotted in Fig.~11 for two
$E_{\nu}$ intervals. Due to a limited resolution of $W$, as well
as the effects of the secondary intranuclear interactions, its
distribution extends to the region below the nucleon mass.
Besides, no peak is seen at  $W \approx m_N$, due to a small
contribution of the identified events of the quasielastic reaction
(1) as compared to the non-identified one. \\

\section{Summary}

The charged current neutrinonuclear reactions at $E_{\nu}$ = 1 - 3
are investigated using the data obtained with SKAT propane-freon
bubble chamber irradiated in the neutrino beam at Serpukhov
accelerator. The reconstructed neutrino spectrum is found to be
consistent with the expected one on the bubble chamber. \\ For the
first time, in this $E_{\nu}$ range detailed experimental data are
obtained on the characteristics of the produced particles. The
$E_{\nu}$ - dependence of the mean multiplicities of protons,
charged pions, $\gamma$- quanta, $\pi^0$ mesons and neutral
strange particles and their multiplicity, momentum and angular
distributions are measured. An indication is obtained, that the
yield of pions in the neutrinonuclear interactions is reduced as
compared to that in $\nu N$ interactions, which can be
qualitatively explaned by their absorption in the nuclear medium.

  {\bf Acknowledgement.} The activity of one of the autors (Zh.K.) is
supported by Cooperation Agreement between DESY and YerPhI signed
on December 6, 2002. The autors from YerPhI (M.A., N.G. and H.G.)
acknowledge the supporting grants of Galust Gulbekian Foundation
and Swiss Fonds "Kidagan".


\begin{figure}[ht]
 \resizebox{1.2\textwidth}{!}{\includegraphics*[bb =80 10 600
510]{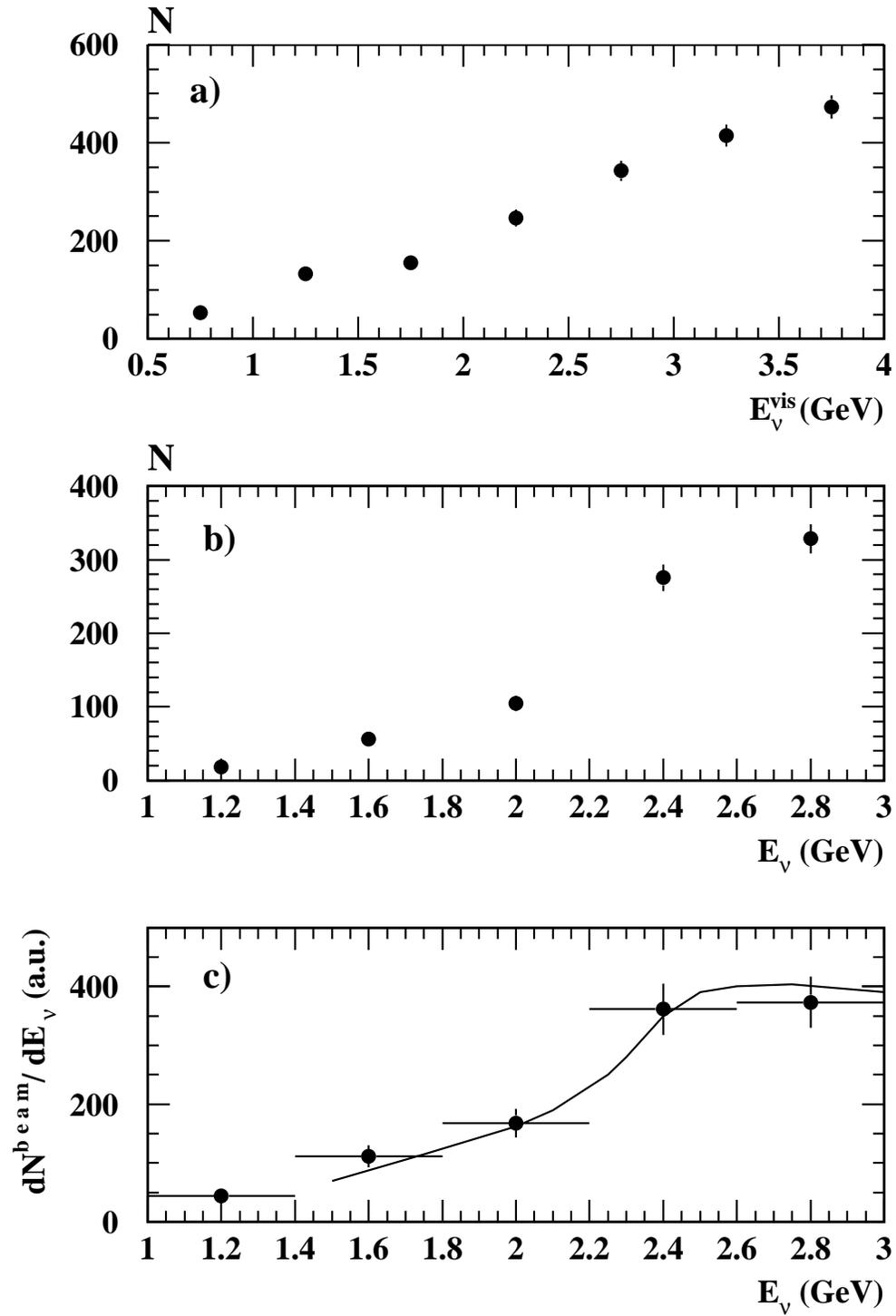}}
 \caption{The distribution on: $E_{\nu}^{vis}$ for selected events (a),
the reconstructed neutrino energy (b), the reconstructed energy of
the beam neutrino (c). The curve is taken from \cite{ref5}.}
\end{figure}

\newpage
\begin{figure}[hb]
\resizebox{0.7 \textwidth}{!}{\includegraphics*[bb=5 10 400 450]
{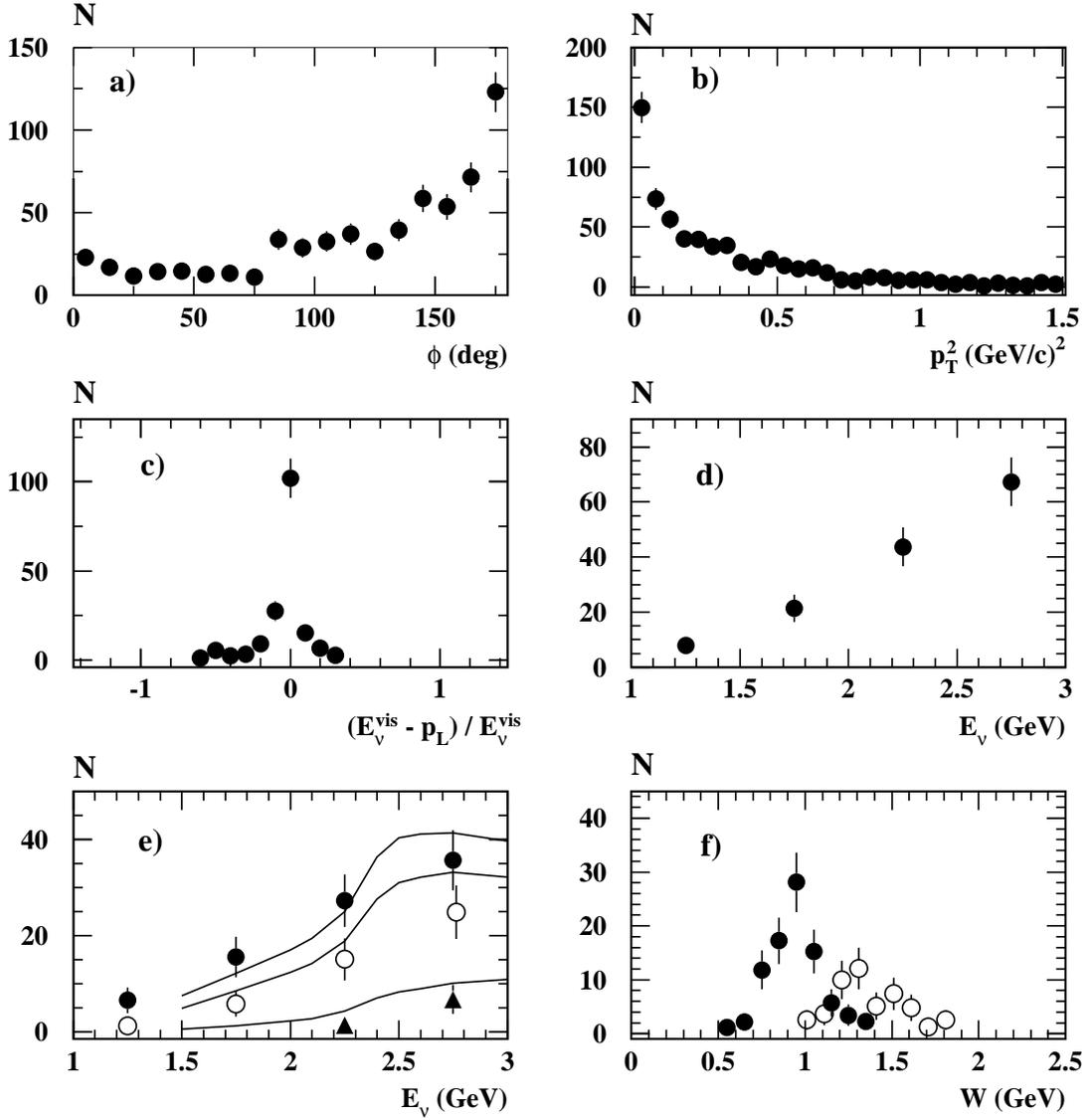}} \caption{The distribution for events-candidates for
exclusive channels on: the azimutal angle $\phi$ between the
momenta of the muon and the hadronic system (a); squared
transverse momentum $p_T^2$ (b); ratio $(E_{\nu}^{vis} -
p_L)/E_{\nu}^{vis}$ for events with $\phi > 155^{\circ}$ and
$p_T^2 <$ 0.15 (GeV$/c)^2$ (c); $E_{\nu}$ for exclusive channels
(1)-(3) (d); $E_{\nu}$ for channels: (1) (full circles), (2)
(empty circles) and (3) (triangles) (e); invariant hadronic mass
$W$ for exclusive channels (1)-(3) (full circles) and the channel
(2) (empty circles) (f). The curves in Fig.2e correspond to the
expected dependences for neutrino interactions with (quasi) free
nucleons (see the text).}
\end{figure}

\newpage
\begin{figure}[h]
\resizebox{1.1 \textwidth}{!}{\includegraphics*[bb=50 10 600
520]{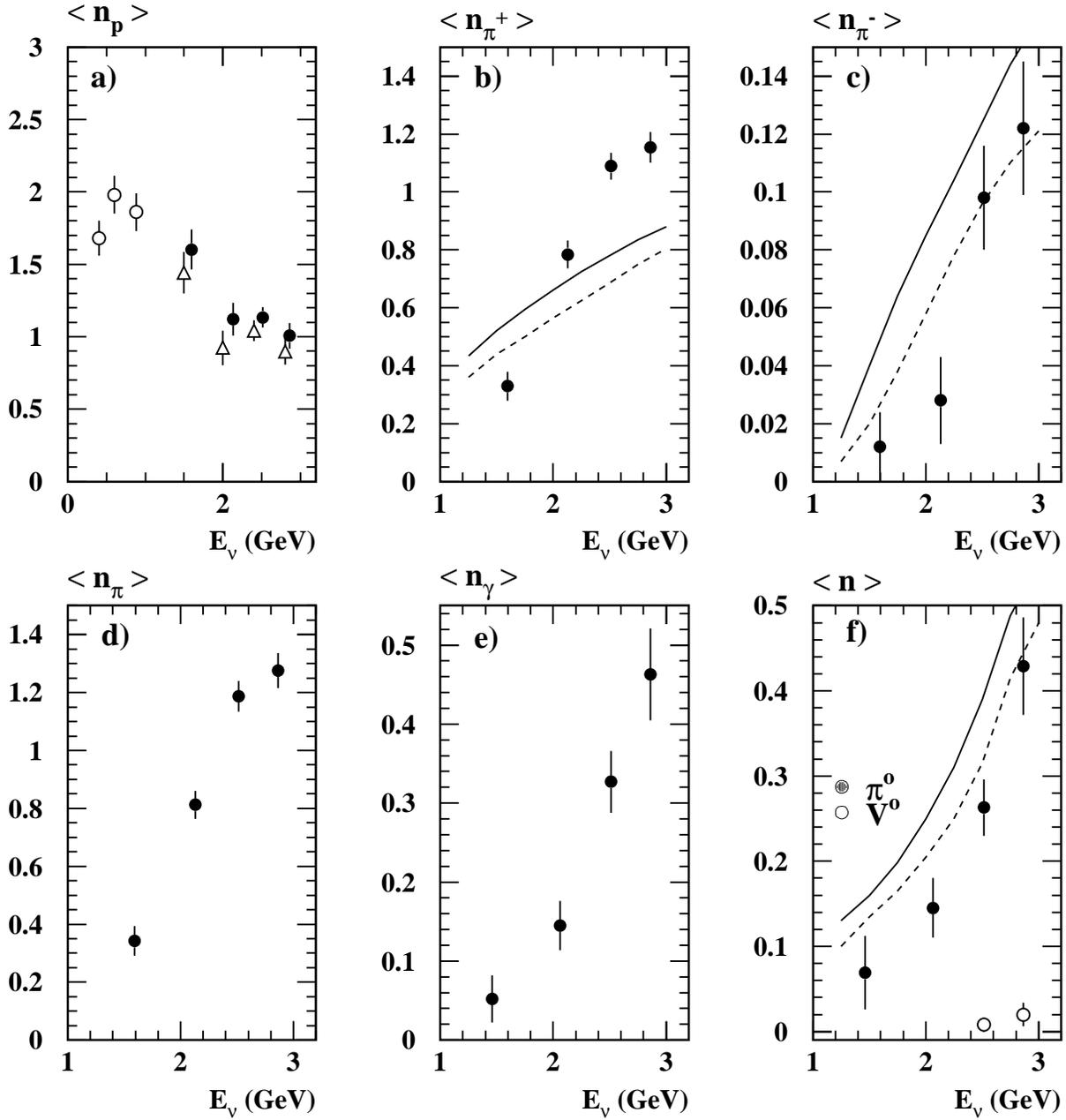}} \caption{The $E_{\nu}$- dependence of the mean
multiplicity of: identified protons (empty circles \cite{ref2} and
triangles) and protons, including 'kinematically'
identified ones (full circles) (a);
${\pi}^+$ mesons (b);
${\pi}^-$ mesons (c); charged pions (d); $\gamma$- quanta (e);
$\pi^0$ mesons (full circles) and $V^0$ (empty circles) (f).
The solid curves are the approximation of the experimental data
on the mean multiplicities of pions in $\nu N$ interactions, while
the dashed curves are those when taking into account the
intranuclear absorption of pions (see the text).}
\end{figure}

\newpage
\begin{figure}[ht]
\resizebox{1.1 \textwidth}{!}{\includegraphics*[bb=50 10 600
520]{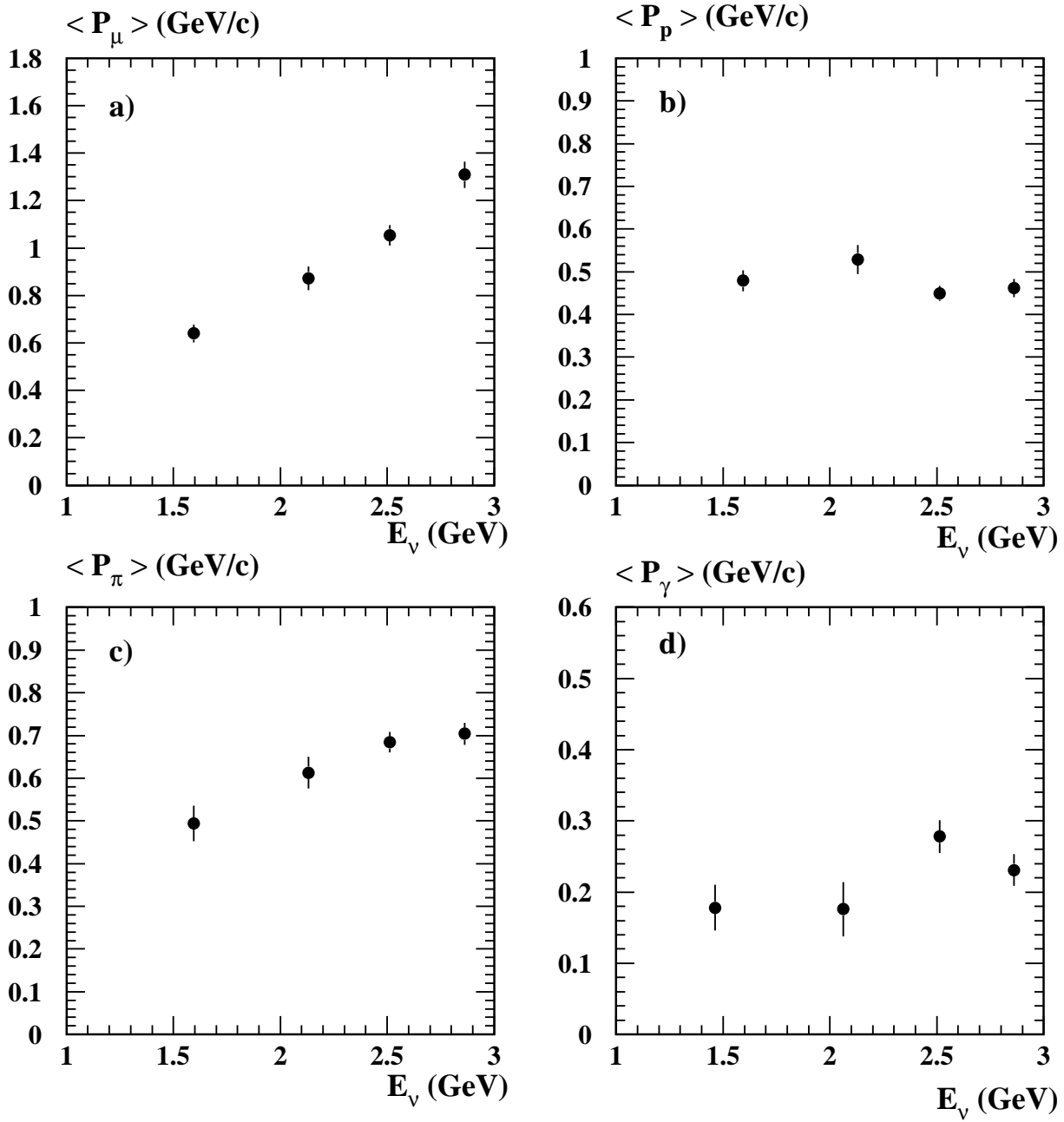}} \caption{The $E_{\nu}$ - dependence of the mean
momentum of: muons (a); protons (b); charged pions (c); $\gamma$-
quanta (d).}
\end{figure}

\newpage
\begin{figure}[ht]
\resizebox{1.1 \textwidth}{!}{\includegraphics*[bb=50 10 600
520]{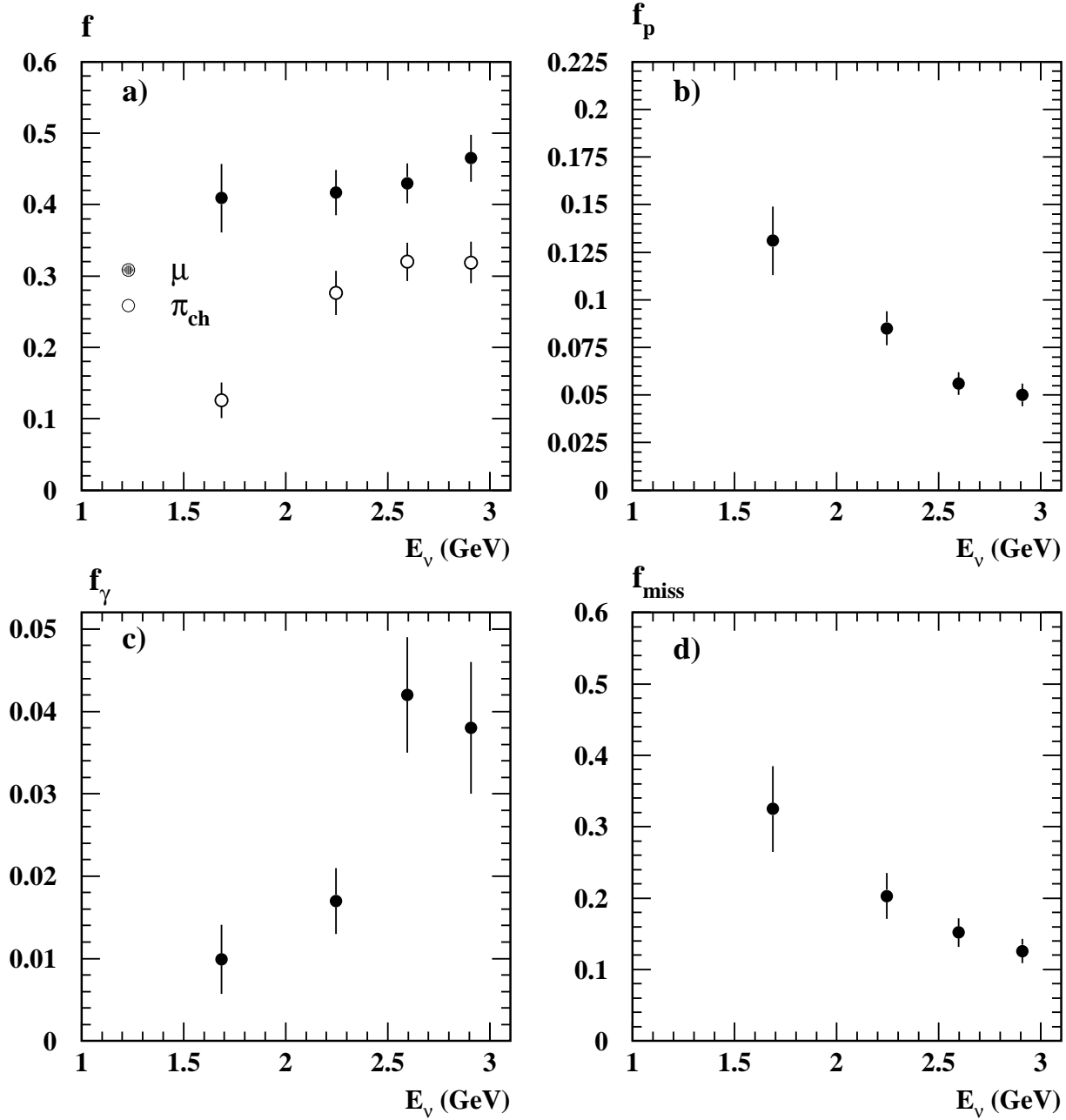}} \caption{The $E_{\nu}$ - dependence of the mean
energy fracton of the neutrino carried by
: muons and charged pions (a); protons (b); $\gamma$- quanta (c);
undetected neutral particles (d).}
\end{figure}

\newpage
\begin{figure}[ht]
\resizebox{1.1 \textwidth}{!}{\includegraphics*[bb=50 10 600
520]{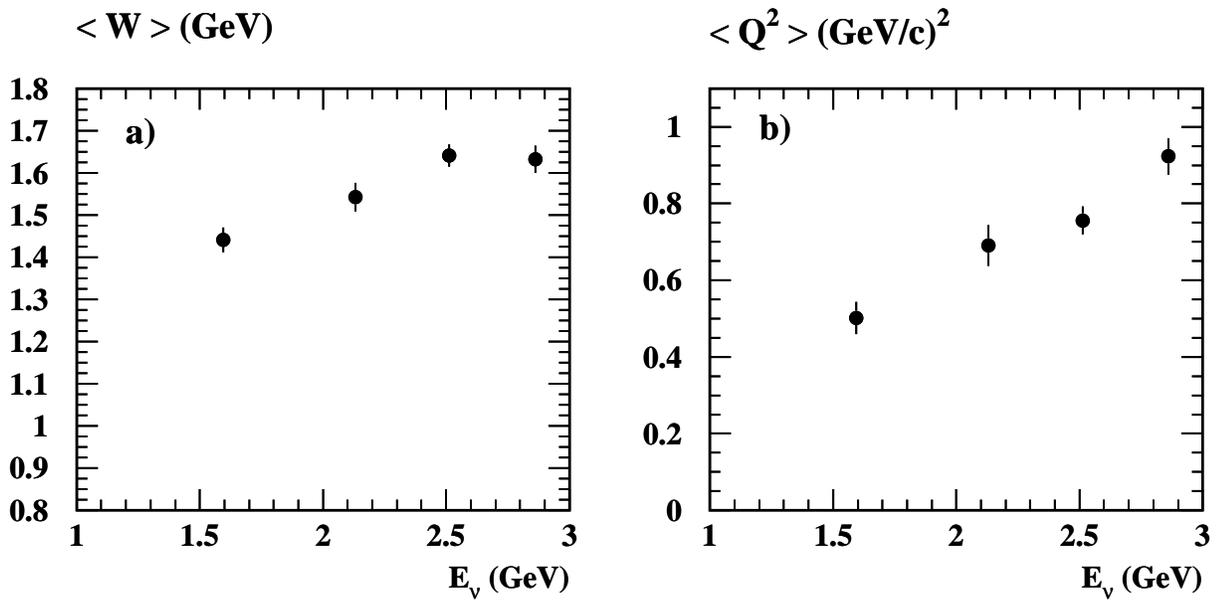}} \caption{The $E_{\nu}$ - dependence of the mean
values of $W$ (a) and $Q^2$ (b)}.
\end{figure}

\newpage
\begin{figure}[ht]
\resizebox{1.1 \textwidth}{!}{\includegraphics*[bb=50 10 600
520]{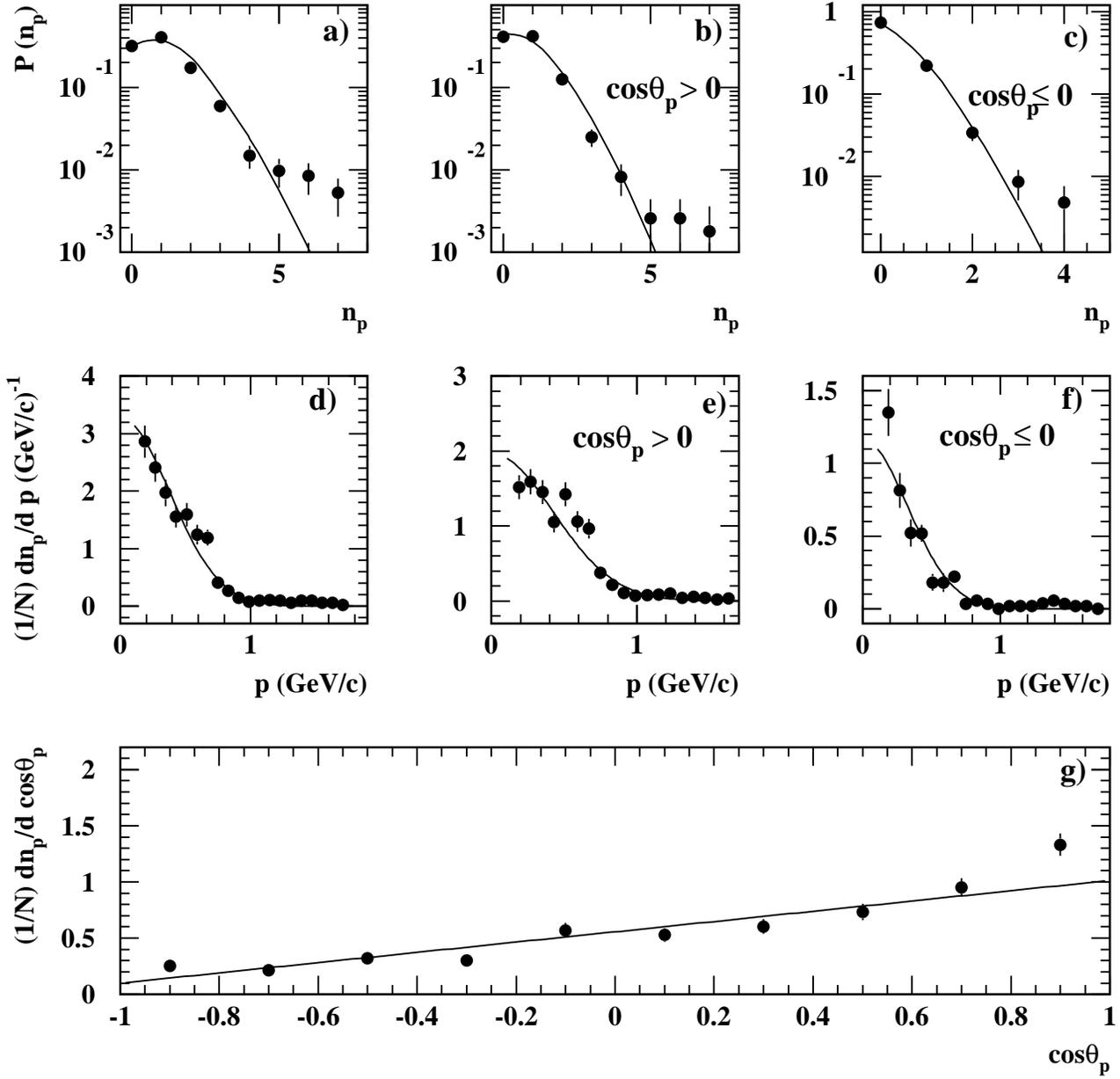}} \caption{The multiplicity and momentum distributions
for protons (a and d); protons with $cos{\theta_p} >$ 0 (b and e)
and $cos{\theta_p} <$ 0 (c and f); the distribution on
$cos{\theta_p}$ (g).}
\end{figure}

\newpage
\begin{figure}[ht]
\resizebox{1.1 \textwidth}{!}{\includegraphics*[bb=50 10 600
520]{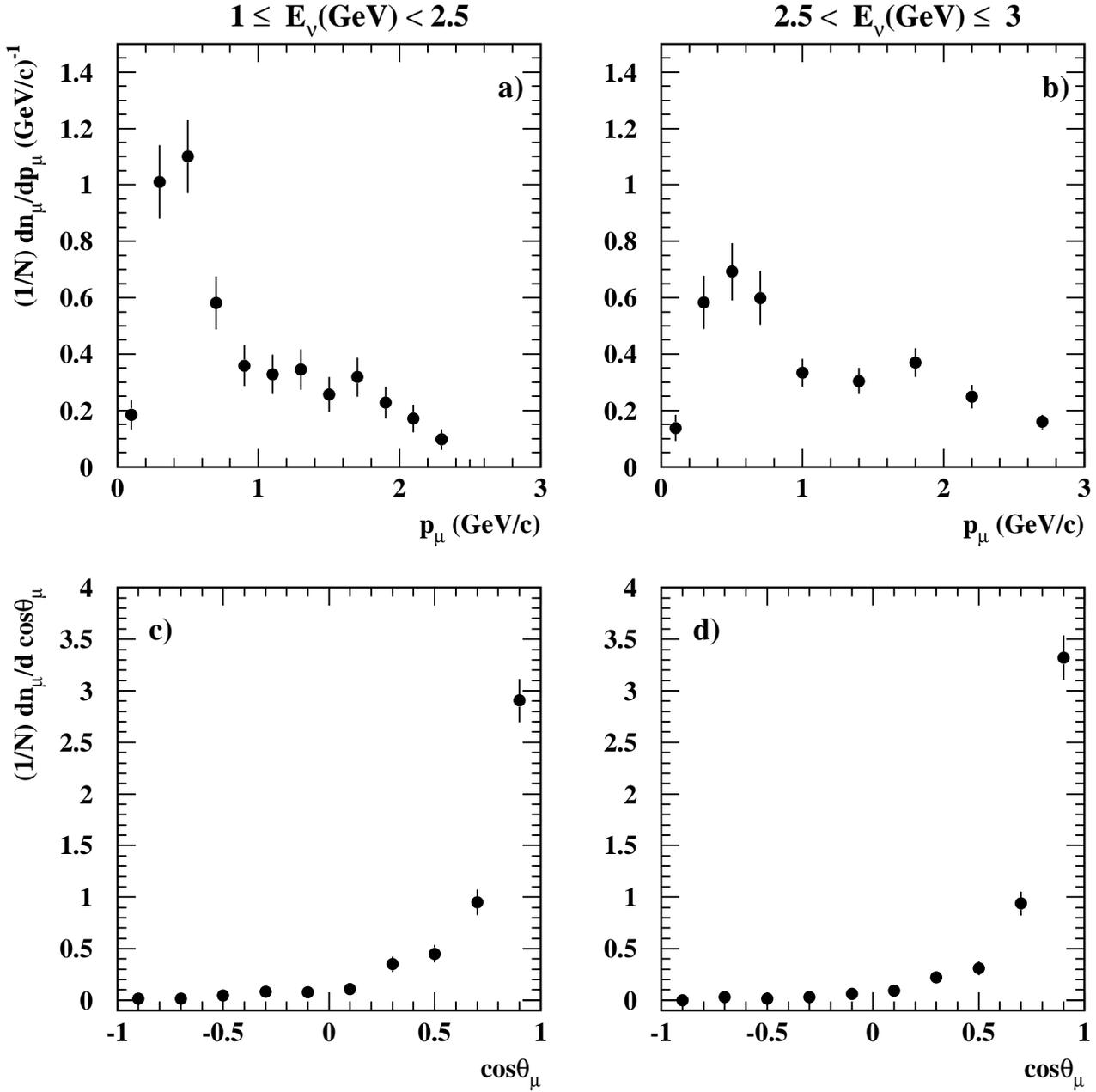}} \caption{The momentum and angular distributions
of muons for two $E_\nu$ intervals.}
\end{figure}

\newpage
\begin{figure}[ht]
\resizebox{1.1 \textwidth}{!}{\includegraphics*[bb=50 10 600
520]{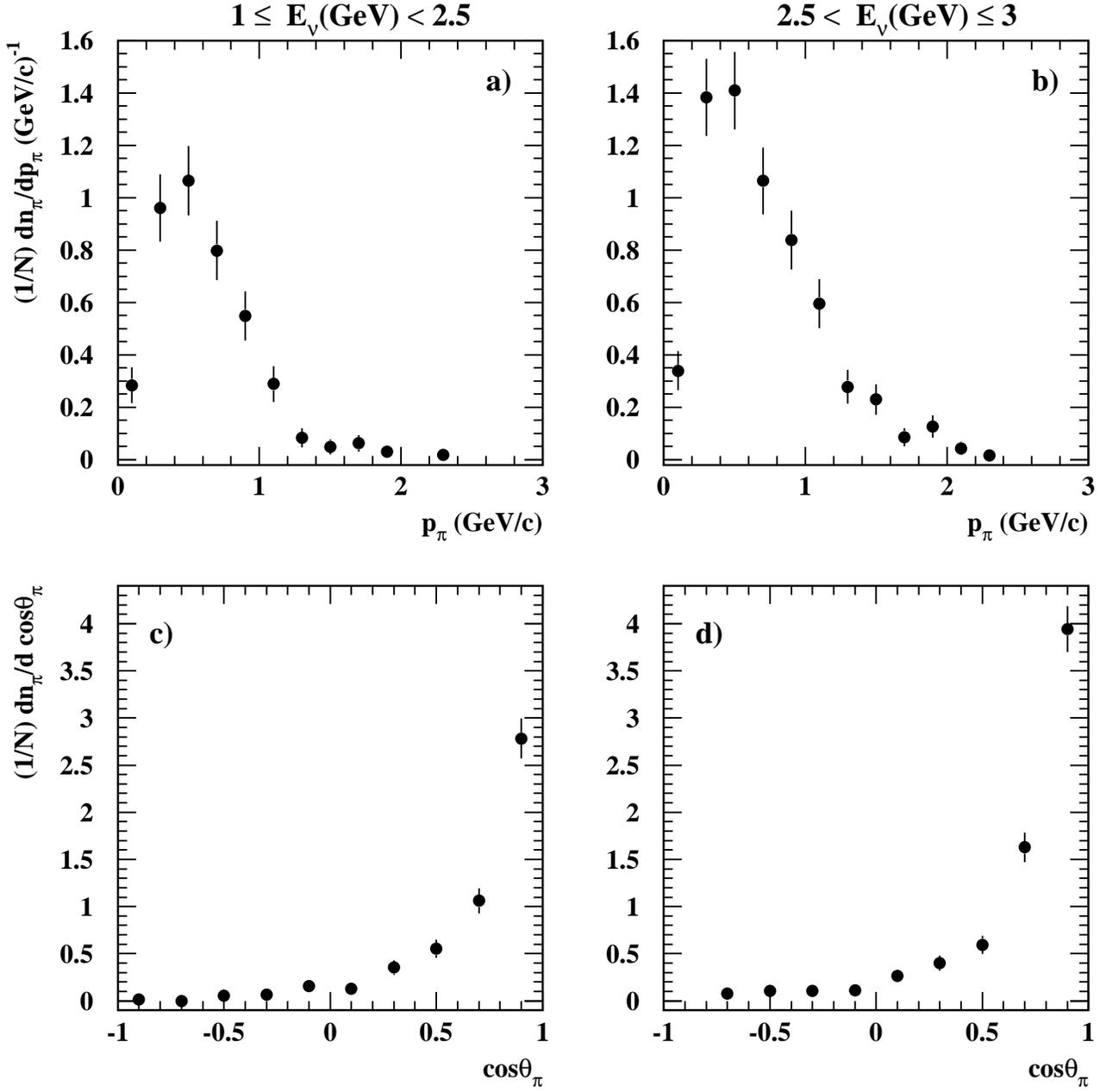}} \caption{The momentum and angular distributions
of charged pions for two $E_\nu$ intervals.}
\end{figure}

\newpage
\begin{figure}[ht]
\resizebox{1.1 \textwidth}{!}{\includegraphics*[bb=50 10 600
520]{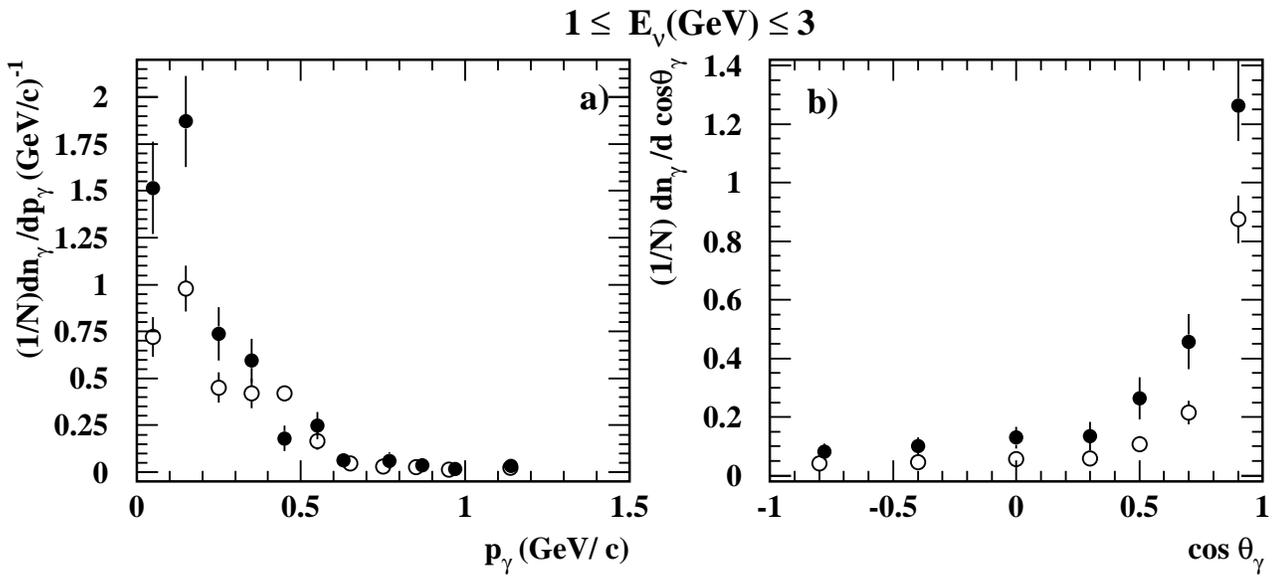}} \caption{The momentum and angular distributions
of $\gamma$- quanta corrected (close circles) and not corrected
(open circles) for their detection efficiency.}
\end{figure}

\newpage
\begin{figure}[ht]
\resizebox{1.1 \textwidth}{!}{\includegraphics*[bb=50 10 600
520]{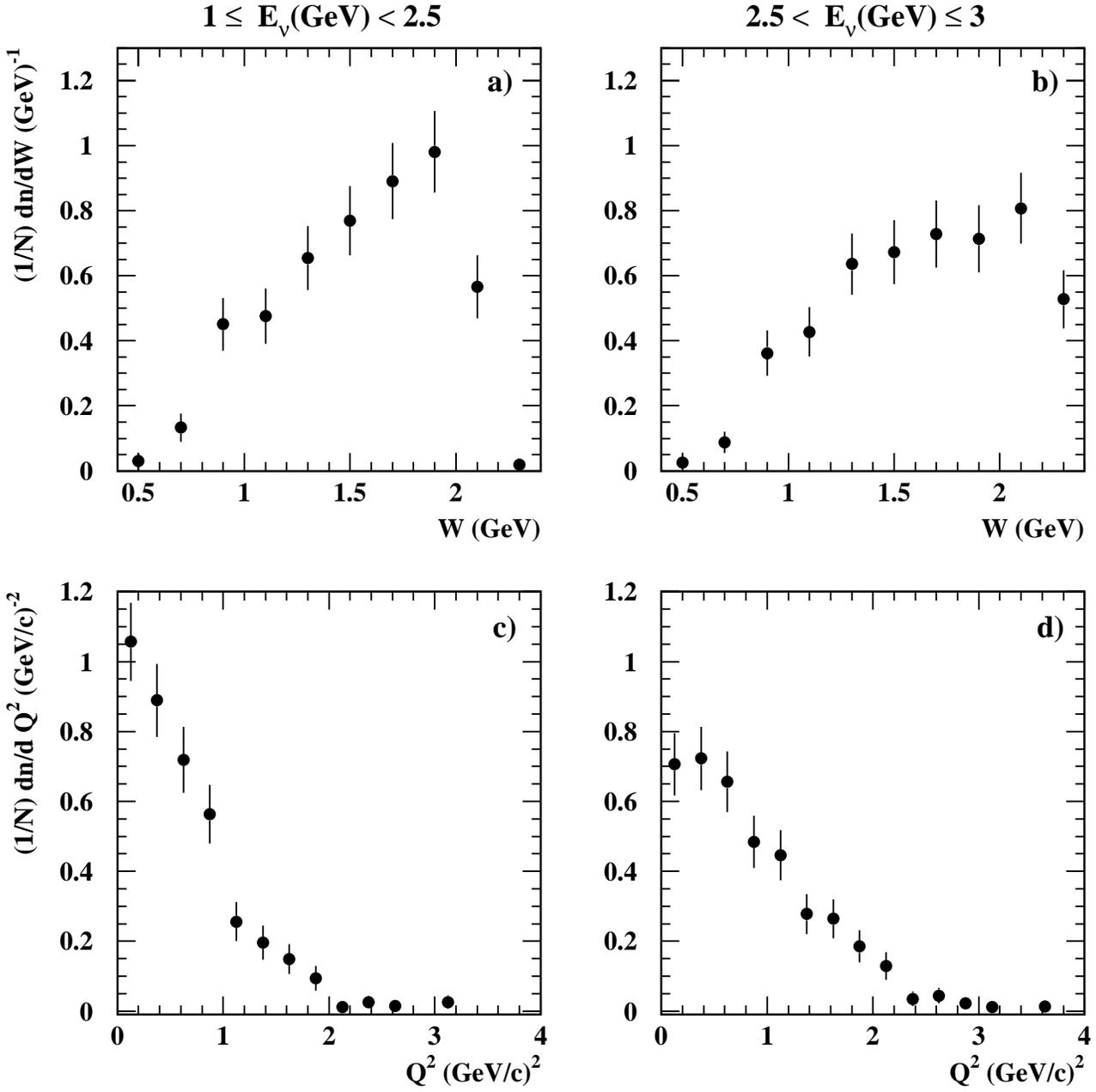}} \caption{The distributions on $W$ (a)
and $Q^2$ (b)}.
\end{figure}

\end{document}